\documentclass[12pt,preprint]{aastex}

\begin{document}
\pagenumbering{arabic}

\title{HOW DO GALAXIES FORM?}

\author{Sidney van den Bergh}
\affil{Dominion Astrophysical Observatory, Herzberg Institute of Astrophysics, National Research
Council of Canada, 5071 West Saanich Road, Victoria, BC, V9E 2E7, Canada}
\email{sidney.vandenbergh@nrc-cnrc.gc.ca}

\begin{abstract}
It is currently believed that galaxies were assembled
via chaotic hierarchical mergers between massive cold 
dark matter halos, in which baryonic star forming matter 
was embedded. One would therefore expect the properties
of individual galaxies to be determined by numerous 
independent factors such as star forming history, merger 
history, mass, angular momentum, size and environment. 
It is therefore surprising to find that galaxies actually appear
to form an (almost) one parameter family in which galaxy
mass is the dominant factor.
\end{abstract}

   New observations often provide a fresh perspective on the
Universe, thus  providing a new window through which to view  
ancient problems. Such is the case for a new survey by  Disney et al.$^{1}$ 
in which they provide a catalog of galaxies 
that were detected during a large blind survey of the sky in the
21-cm line of neutral hydrogen. For each of the $\sim$200 objects in 
their catalog the authors measured the total hydrogen mass, 
the width of the 21-cm line, the redshift and inclination to
the line of sight, two radii and the optical luminosity in four 
different color bands. The HI-selected galaxies appear to
have colors made up of two components; a `rogue' color,
which probably correlates with recent star forming history,
and a systematic component related to the mean age of
the stellar population in that galaxy. [Both rogue color and
the Hubble morphological types of galaxies are strongly
correlated with environment].  Since the nature of a
galaxy's color is complex Disney et al. initially neglect it 
and concentrate on the other factors that describe the
galaxies in their sample. From a principal component analysis
of their data they find that the six components that they use
to describe the galaxies in their sample (including the 
systematic color) are all correlated with each other and with 
a single principal component (mass). In other words galaxies 
appear to constitute a single `Fundamental Line'  in parameter 
space.  Disney et al. argue that such simplicity is difficult to 
understand within the framework of the hierarchical galaxy 
merging paradigm. This is so because  it is expected that the 
evolution of individual galaxies will depend on (1) their initial 
spins (2) on a the details of their chaotic merger history and (3)  
on the density, temperature and turbulence of the environment 
in which a galaxy is assembled. 

The work by Disney et al. follows
 closely in the footsteps of an earlier study by Gavazzi et al $^{2}$ in 
which it was shown that the radii of galaxies, measured at a blue
 surface brightness of 25.0 magnitudes per square arcsecond, 
were very closely correlated galaxy luminosity. From this 
observation Gavazzi et al. concluded that, to first order, galaxy
 evolution can be described by a single parameter  - the initial 
mass. In particular they pointed out that the Hubble 
morphological types of a galaxies exhibits little dependence 
on their structural properties. They therefore concluded that 
such factors as environment and initial angular momentum 
induce only second order effects on the overall evolution of 
galaxies. Surprisingly Gavazzi et al. note that the correlation 
between the radii and the luminosities of galaxies appears to 
differ little, if at all, between relatively isolated field galaxies 
and galaxies that are located in rich clusters. They conclude 
from these results that gravitational interactions between 
galaxies may have played a lesser role than previously believed.
A similar result was obtained by Girardi et al. $^{3}$ who divided
 observations of galaxies in the Virgo cluster into sub-regions: 
an inner region  within R = 0.5 Mpc, an intermediate shell with 
 0.5 $<$ R $<$ 1.0 Mpc, and an outer zone with R $>$ 1.0 Mpc from 
the cluster center. For the galaxies in each of these three regions 
of the Virgo cluster Girardi et al. obtained  similar luminosity-
radius relations. To check on the conclusion that the relation
between the luminosities of galaxies and their radii  are not 
strongly correlated with environment van den Bergh $^{4}$ has 
recently compared the the radii and the luminosities of the 80 
brightest galaxies within a distance of 10 Mpc (33 million light-
years) of the Sun. These data  show that nearby galaxies seem 
to exhibit the same relationship between  radius and luminosity 
as do the galaxies in great clusters such as Virgo and Coma. 
Furthermore these nearby galaxies also appear to exhibit no 
obvious correlations between their luminosity-radius relation and 
their present environment. In particular no dependence is found 
on the local mass density as defined by Karachentsev \& 
Malakov $^{5}$.
 
   The mounting volume of evidence discussed above suggests 
that galaxies constitute an (almost) one parameter family based 
on mass, with little or no indication of a significant dependence 
on environmental density. This conclusion poses  a number of 
significant challenges to the prevailing hierarchical merging 
paradigm. One is reminded of the saying by Danish poet 
Piet Hein that:

 Problems worthy of attack prove their worth by hitting back.

\end{document}